\documentclass[prb, aps, reprint, superscriptaddress, floatfix]{revtex4-2}
\usepackage{amsmath}
\usepackage{amssymb}
\usepackage{bm}
\usepackage{graphicx}
\usepackage[colorlinks=true, linkcolor=blue, citecolor=blue, urlcolor=blue]{hyperref}

\graphicspath{{./figures/}}

\usepackage[svgnames]{xcolor}

\begin{document}
\title{Magnetic skyrmions embedded in a vortex}

\date{\today}

\author{Haobing Zhang}
\affiliation{Institutes of Physical Science and Information Technology, Anhui University, Hefei 230601, China}

\author{Xintao Fan}
\email{fanxt8@163.com}
\affiliation{School of Materials Science and Engineering, Anhui University, Hefei 230601, China}

\author{Weiwei Wang}
\email{wangweiwei@ahu.edu.cn}
\affiliation{Institutes of Physical Science and Information Technology, Anhui University, Hefei 230601, China}

\begin{abstract}
Magnetic vortices and skyrmions represent two fundamental classes of topological spin textures in ferromagnetic systems, 
distinguished by their unique stabilization mechanisms and degrees of freedom. Vortices, characterized by circular in-plane magnetization (chirality) 
and out-of-plane core polarization, naturally arise in confined geometries due to the interplay between exchange and dipolar interactions. 
In contrast, skyrmions typically require the Dzyaloshinskii-Moriya interaction for stabilization and exhibit fixed chirality-polarity relationships. 
Through micromagnetic simulations, we reveal that these seemingly distinct topological states can coexist, forming a novel composite state termed 
the \textit{n}-skyrmion vortex, which represents a skyrmion-embedded vortex state. These composite states possess quantized topological charges 
$Q$ that follow the relation $Q_{\text{total}} = Q_{\text{vortex}} + nQ_{\text{skyrmion}}$, where $n$ denotes the number of embedded skyrmions. 
Similar to vortices, these states exhibit independent chirality and polarity and are energetically degenerate.
\end{abstract}

\maketitle

\section{Introduction}

Magnetic vortices represent a fascinating class of topological spin textures that emerge in soft magnetic materials, 
particularly in geometrically confined systems such as thin films and nanodots~\cite{shinjo2000, wachowiak2002}. 
These equilibrium configurations arise from the delicate balance between exchange interactions and magnetostatic energies, 
with their stability spanning an impressive size range from sub-100 nm nanostructures to micron-scale systems~\cite{Guslienko2006c, Chung2010}. 
The fundamental characteristics of a magnetic vortex are defined by two independent parameters: \textit{chirality} (clockwise 
or counterclockwise in-plane magnetization circulation) and \textit{polarity} (upward or downward out-of-plane core magnetization)~\cite{jaafar2010}. 
This dual degree of freedom enables four distinct magnetic states through various chirality-polarity combinations, 
making vortices particularly attractive for multistate memory applications~\cite{cambel2011a}.

In contrast to these naturally occurring vortices, magnetic skyrmions -- topologically protected quasiparticles with quantized winding numbers -- 
require the presence of Dzyaloshinskii-Moriya interaction (DMI) for their stabilization~\cite{Dzyaloshinskii1958, Moriya1960}. 
The DMI symmetry dictates the skyrmion type: interfacial DMI in asymmetric multilayers (e.g., Pt/Co/Ir) generates Néel-type skyrmions 
with radial spin modulations, while bulk DMI in non-centrosymmetric crystals (e.g., MnSi) produces Bloch-type skyrmions exhibiting 
tangential spin rotations~\cite{bogdanov2001, rohart2013}. Unlike vortices, skyrmions exhibit an intrinsic locking mechanism 
between their chirality and polarity, effectively reducing their configurational degrees of freedom~\cite{moreau-luchaire2016, Zhang2015}. 
This fundamental difference in topological freedom has significant implications for their dynamical behavior and technological applications~\cite{Fert2017}.

Recent advances have unveiled rich interactions between these topological structures, particularly through the interplay of DMI
and magnetostatic effects in confined geometries. Theoretical studies have demonstrated that DMI can transform conventional 
vortices into chiral spin textures, including radial vortex states with enhanced spin modulations~\cite{Siracusano2016b}. 
This transformation is driven by the competition between DMI-induced chirality and the magnetostatic energy of the vortex core. 
Moreover, ultrafast magnetization dynamics have revealed that sub-nanosecond magnetic field pulses can precisely control vortex 
chirality transitions via excitation of gyrotropic core modes~\cite{vanwaeyenberge2006a, hertel2007a}. Moreover, microwave fields 
can selectively manipulate skyrmion core polarization, offering an alternative approach for topological state switching~\cite{Zhang2015}. 

In this study, we employ micromagnetic simulations to demonstrate that competing energy scales -- particularly the interplay among 
DMI strength, magnetostatic interactions, and geometrical confinement -- stabilize novel topological hybrid states, which 
we designate as \textit{n-Skyrmion Vortex}. By systematically exploring the parameter space, including external fields, anisotropy constants and system radii, 
we identify distinct phase regimes where the \textit{n-Skyrmion Vortex} emerges as a stable configuration. These hybrid states exhibit quantized 
topological charges given by $Q_{\text{total}} = Q_{\text{vortex}} + n\,Q_{\text{skyrmion}}$, with \(n\) representing the number of embedded skyrmions.

\begin{figure*}[tbhp]
  \centering
  \includegraphics[width=0.98\textwidth]{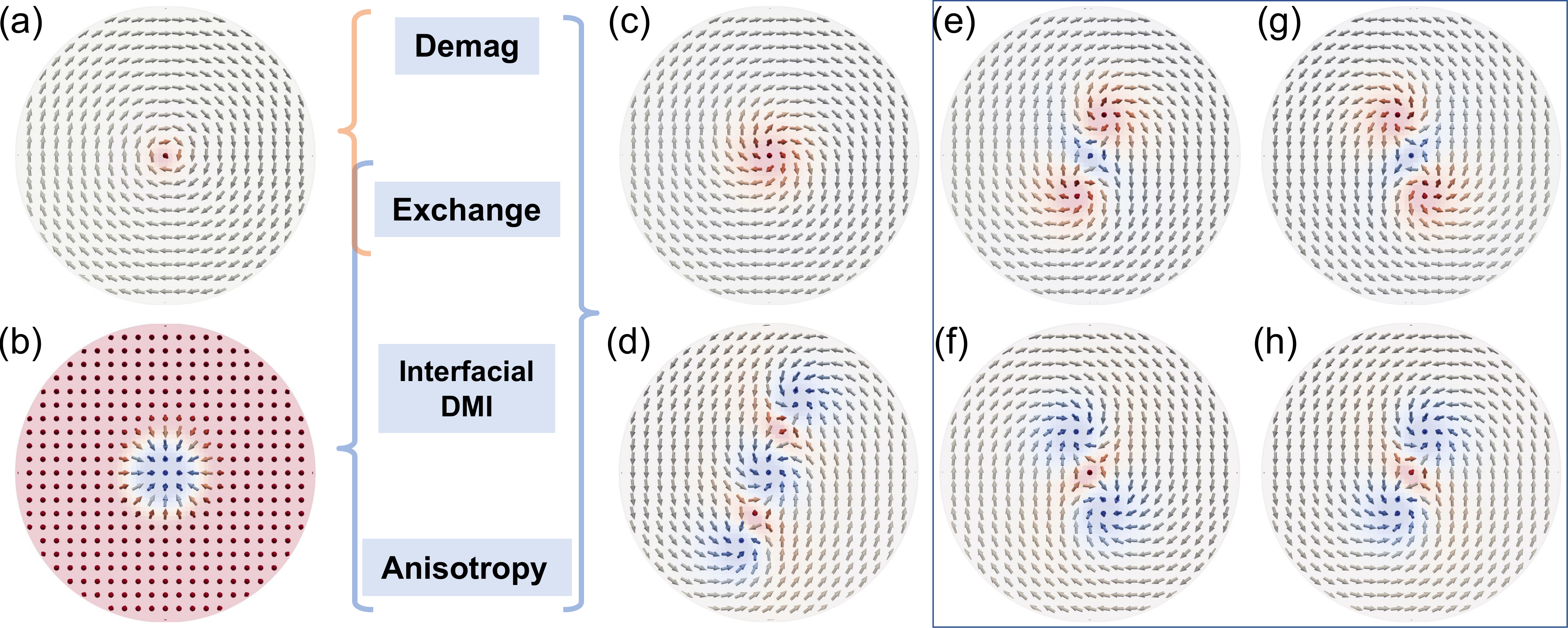}
  \caption{Topological spin textures in a 400\,nm diameter ferromagnetic disk (only the central 130\,nm region is shown for clarity). 
  (a) Conventional vortex state ($Q = +0.5$) stabilized by exchange and dipolar interactions ($D = 0$ and $K = 0$). 
  (b) Isolated Néel-type skyrmion ($Q = -1$) stabilized by anisotropy, exchange, and interfacial DMI in the absence of dipolar interactions. 
  (c) DMI-modified vortex state ($Q = -0.5$) exhibiting an expanded core region due to interfacial DMI ($D = 2 \times 10^{-3}$\,J/m$^2$). 
  (d) 2-skyrmion vortex complex ($Q = -2.5$) containing two embedded skyrmions. 
  (e) 1-skyrmion vortex complex ($Q = +1.5$) formed by embedding a single skyrmion within a vortex background. 
  Panels (e-h) display four degenerate configurations of the 1-skyrmion vortex complex, demonstrating independent control of chirality and polarity: 
  (e) clockwise chirality with $p = +1$, (g) counterclockwise chirality with $p = +1$, (h) clockwise chirality with $p = -1$, and (i) counterclockwise chirality with $p = -1$. 
  Configurations (c-h) were obtained with a uniaxial anisotropy of $K = 4 \times 10^5$\,J/m$^3$. Zero external field $H_z = 0$\,mT is used.}
  \label{fig1}
\end{figure*}

\section{Model and Simulations}
We consider a circular thin film system described by the following energy contributions: exchange interaction, 
Zeeman energy, uniaxial anisotropy, interfacial Dzyaloshinskii-Moriya interaction (DMI), and dipolar interactions. 
The total free energy of the system is given by:
\begin{equation}\label{eq_energy}
    E = \int \left[ A (\nabla \mathbf{m})^2 - K m_z^2 + E_\mathrm{ze} + E_\mathrm{dmi} + E_d \right] dV
\end{equation}
where $\mathbf{m} = \mathbf{M}/M_s$ is the normalized magnetization vector, $A$ is the exchange stiffness constant, 
$K$ is the uniaxial anisotropy constant, $E_\mathrm{dmi}$ represents the DMI energy density, and $E_d$ denotes the 
demagnetization energy density. The Zeeman energy $ E_\mathrm{ze}  = - \mu_0 M_s H_z m_z$ for the external magnetic field along 
the $z$-axis (perpendicular to the film plane).

To determine the stable magnetic configurations, we employ two complementary approaches: 1) Energy minimization using a gradient descent 
method with adaptive step sizes determined by the Barzilai-Borwein algorithm. 2) Dynamic relaxation through numerical integration of the 
Landau-Lifshitz-Gilbert (LLG) equation:
\begin{equation}\label{eq_LLG}
    \frac{\partial \mathbf{m}}{\partial t} = -\gamma \mathbf{m} \times \mathbf{H}_\mathrm{eff} + \alpha \mathbf{m} \times \frac{\partial \mathbf{m}}{\partial t}
\end{equation}
Here, $\gamma = 2.21 \times 10^5~\mathrm{m/(A\cdot s)}$ is the gyromagnetic ratio, 
$\mathbf{H}_\mathrm{eff} = -(1/\mu_0 M_s) \cdot \delta E/\delta \mathbf{m}$ is the effective field, 
and $\alpha$ is the Gilbert damping parameter. The energy terms in Eq.~(\ref{eq_energy}) are computed using finite difference methods.
All micromagnetic simulations were performed using the GPU-accelerated finite difference micromagnetic 
package~\texttt{MicroMagnetic.jl}~\cite{Wang2024a}.
The simulation parameters of CoFeB were adopted from Ref.~\cite{Siracusano2016b, Mustaghfiroh2019}: 
the saturation magnetization \( M_s = 1.1 \times 10^6 \) A/m, 
the exchange constant \( A = 1.3 \times 10^{-11} \) J/m, the interfacial DMI constant 
\( D = 2 \times 10^{-3} \) J/m$^2$. The thickness of the film is fixed as $2$ nm
and the cellsize is chosen to be $2\times 2\times 2~\mathrm{nm}^3$.

\section{Results}

Magnetic textures such as skyrmions and vortices can be characterized by a non-zero topological charge \( Q \) (skyrmion number). 
The topological charge \( Q \) is calculated as the integral of the topological density over the two-dimensional plane:
\begin{equation}
Q = \frac{1}{4\pi} \int \mathbf{m} \cdot \left( \frac{\partial \mathbf{m}}{\partial x} \times \frac{\partial \mathbf{m}}{\partial y} \right) dx\,dy,
\end{equation}
where $\mathbf{m}$ is the unit vector of the magnetization direction. 
This quantity measures the number of times the magnetization vector wraps around the unit sphere \citep{Nagaosa2013}.
Magnetic skyrmions typically exhibit quantized topological charges \( Q = \pm 1 \), 
while vortices are characterized by half-integer values \( Q = \pm 1/2 \).

In magnetic disks, the competition between dipolar and exchange interactions can give rise to vortex structures. 
Figure~\ref{fig1}(a) illustrates a vortex observed in a circular disk with a radius of 200\,nm. To clearly reveal the underlying magnetic configuration, 
only the central region with a radius of 50\,nm is shown, thereby highlighting the clockwise rotation of the vortex. 
In the presence of exchange interactions, interfacial DMI, and anisotropy, a stable N\'eel-type skyrmion is realized, 
as depicted in Fig.~\ref{fig1}(b). Interestingly, when these four interactions are combined, a distinct magnetic configuration emerges, 
termed the \textit{DMI-modified vortex} or \textit{DMI vortex} for short, see Fig.~\ref{fig1}(c). The incorporation of DMI leads to a less sharp vortex core.
Notably, the magnetic structure in Fig.~\ref{fig1}(c) resembles a magnetic meron~\cite{Lin2015,Yu2018,Xia2022} and 
is achieved for a DMI strength of $D = 2 \times 10^{-3}\,\mathrm{J/m^2}$.
We designate this state as the DMI vortex for two reasons: (i) it can be derived from the standard vortex structure by introducing interfacial DMI, 
and (ii) its chirality and polarity remain independent, resulting in energetically degenerate states analogous to those of the standard vortex. 
The presence of nonzero DMI induces a slight twisting of the magnetization near the edge. Consequently, the skyrmion number is evaluated 
within a circular region of radius 100\,nm, yielding a value of \(Q = +0.5\).

Figure~\ref{fig1}(e) presents a novel magnetic configuration--the \emph{skyrmion-embedded vortex} state--which we refer to as the 1-skyrmion vortex. 
This state exhibits a total topological charge of \(Q = -1.5\), arising from the sum of the skyrmion’s topological charge (\(Q = -1\)) and that of the vortex (\(Q = -0.5\)). 
Notably, despite the presence of DMI, the chirality and polarity in the 1-skyrmion vortex remain independent. Consequently, there exist four energetically degenerate states, 
corresponding to two possible chiralities (clockwise and counterclockwise) and two polarities (\(p = +1\) and \(p = -1\)), as illustrated in Figs.~\ref{fig1}(e)–\ref{fig1}(h). 
Similarly, a vortex can host two skyrmions, forming a 2-skyrmion vortex configuration, as shown in Fig.~\ref{fig1}(d). 
In general, we denote these types of structures as \(n\)-skyrmion vortices.

\begin{figure}[tbhp]
  \centering
  \includegraphics[width=0.48\textwidth]{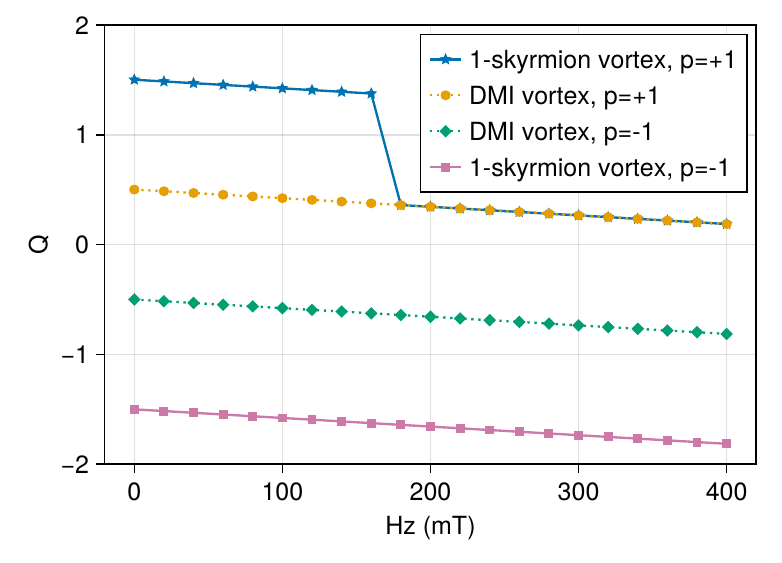}
  \caption{Topological charge evolution under external magnetic fields. Circle/diamond markers: DMI vortex with p=+1/-1 polarity. 
  Rectangle/star markers: 1-skyrmion vortex with p=-1/+1 polarity.}
  \label{fig2}
\end{figure}

A related composite magnetic configuration is known as the \emph{skyrmion bag}~\cite{Foster2019,Kind2021}. 
While both the skyrmion bag and the \(n\)-skyrmion vortex involve the coexistence of topologically nontrivial spin textures, 
the skyrmion bag typically refers to a confined collection of skyrmions within a closed domain wall~\cite{Rybakov2019}, 
whereas the \(n\)-skyrmion vortex describes a vortex state with one or more embedded skyrmions. 
Moreover, the $n$-skyrmion vortex exhibits half-integer topological charges, contrasting with the integer quantization of skyrmion bags.
This fractionalization originates from the vortex background's $Q = \pm 0.5$ contribution.

Figure~\ref{fig2} illustrates the evolution of both the DMI vortex and the 1-skyrmion vortex under an applied external magnetic field. 
As the magnetic field increases, the topological charge \(Q\) of these structures gradually decreases regardless of the polarity. 
For a DMI vortex with \(p = +1\), the zero-field state exhibits \(Q = 0.5\); however, the external field tilts the magnetization vectors 
toward the \(z\)-axis, thereby reducing the number of vectors that wrap the unit sphere and decreasing \(Q\) (as indicated by the circle markers). 
Conversely, for a DMI vortex with \(p = -1\), where the core magnetization points downward, the tilting effect induces additional wrapping 
of the magnetization vectors around the unit sphere. Consequently, although the absolute value of \(Q\) increases, 
the net \(Q\) still decreases from its zero-field value of \(-0.5\) (as shown by the diamond markers). 
The 1-skyrmion vortex follows a similar trend; in particular, the \(p = -1\) case mirrors the behavior of the DMI vortex (see the rectangle markers). 
Notably, when the external field reaches \(H_z = 180\,\mathrm{mT}\), the \(p = +1\) 1-skyrmion vortex converges to the same \(Q\) as the DMI vortex 
(indicated by the star markers), signaling a transition from the 1-skyrmion vortex to the DMI vortex state. 
This observation is consistent with the well-established fact that an isolated skyrmion attains an energetically stable configuration 
only when its core magnetization is oriented antiparallel to the external field.

\begin{figure}[tbhp]
  \includegraphics[width=0.42\textwidth]{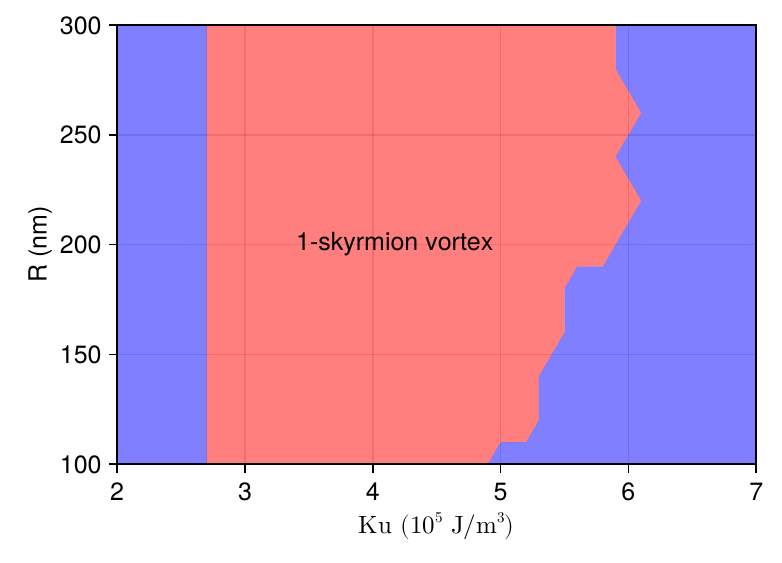}
 \caption{ The phase diagram of anisotropy constant \( K_u \) versus the radius $R$ of the disk, where red and blue regions denote the presence 
 and absence of 1-skyrmion vortex, respectively.}
\label{fig3}
\end{figure}

Moreover, the evolution of the topological charge with the external magnetic field reveals that the \(Q\) curves for different states are parallel, 
with a constant offset of unity between them. For instance, the \(p = -1\) DMI vortex possesses a topological charge exactly one unit larger than 
that of the \(p = -1\) 1-skyrmion vortex. This quantized difference confirms that the 1-skyrmion vortex can be interpreted as a DMI vortex with 
an additional skyrmion embedded within it, thereby affirming the topological charge relation $Q_{\text{total}} = Q_{\text{vortex}} + nQ_{\text{skyrmion}}$ 
with $n$ the number of embedded skyrmions.

The geometry of the disk plays a decisive role in determining the magnetic structure. For example, in a disk with a radius of 125\,nm, 
under the influence of DMI, a radical vortex can be stabilized~\cite{Siracusano2016b}. In our study, a disk with a radius of 200\,nm is employed, 
which, under identical DMI conditions, supports the formation of both the DMI vortex and \(n\)-skyrmion vortex states. 
The increased disk radius results in a higher demagnetization energy, rendering the radical vortex structure less favorable for energy minimization. 
Additionally, magnetic anisotropy can stabilize skyrmions~\cite{Banerjee2014, Zhang2015}, thereby influencing the overall magnetic configuration. 
Figure~\ref{fig3} presents the phase diagram of the 1-skyrmion vortex as a function of the disk radius and the anisotropy constant \(K_u\), 
with the red region delineating the stability window for the 1-skyrmion vortex. It is evident that when \(K_u\) is below approximately \(2.8 \times 10^5\) J/m\(^3\), 
the 1-skyrmion vortex cannot be stabilized. Moreover, the upper bound of \(K_u\) necessary for stabilizing the 1-skyrmion vortex depends on the disk radius, 
reflecting the influence of the effective demagnetization energy in these systems.

\begin{figure}[tbhp]
  \includegraphics[width=0.45\textwidth]{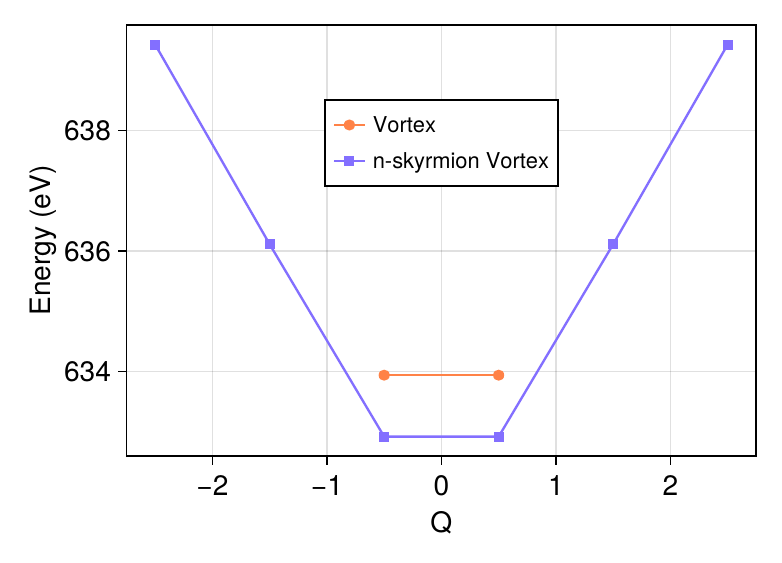}
 \caption{The energy of vortex and $n$-skyrmion vortex as function of topological charge $Q$ with $K=4 \times 10^5$ J/m$^3$. 
 The energy of vortex is calculated with $D=0$.}
\label{fig4}
\end{figure}

Figure~\ref{fig4} shows the dependence of the energy on the topological charge \(Q\) for various \(n\)-skyrmion vortex states. 
For comparison, the energy of the standard vortex (calculated with \(D = 0\)) is also displayed. It is observed that the energy of the DMI vortex formed 
at \(D = 2\times10^{-3}\,\mathrm{J/m^2}\) is slightly lower than that of the standard vortex. Furthermore, for the \(n\)-skyrmion vortices, 
the energy increases with the number of embedded skyrmions. The monotonic increase in energy with \( |Q| \) further 
confirms that the system prefers configurations with smaller topological charges under the given conditions. 
This behavior is consistent with theoretical predictions for skyrmion bag in the presence of uniaxial anisotropy~\cite{Rybakov2019}.

It is worth noting that although our investigation is based on a circular disk geometry, similar \(n\)-skyrmion vortex structures can 
also be realized in square thin films. In circular disks, the \(n\)-skyrmion vortex does not exhibit any preferential orientation 
due to the rotational symmetry of the system. In contrast, the broken rotational symmetry in square thin film systems 
induces a distinct directional preference in the orientation of the \(n\)-skyrmion vortex.

\section{Conclusion}

In summary, our micromagnetic simulations reveal that the interplay among exchange, dipolar, and interfacial Dzyaloshinskii–Moriya interactions
-- combined with magnetic anisotropy and geometrical confinement -- can stabilize novel composite topological states in ferromagnetic systems. 
In particular, we have identified the emergence of \textit{n}-skyrmion vortices, where one or more skyrmions are embedded within a vortex structure, 
resulting in a half-integer quantized topological charge. These hybrid states exhibit independent chirality and polarity.
Our findings not only bridge the gap between conventional vortices and skyrmions but also pave the way for designing tunable topological magnetic textures 
with potential applications in spintronic devices.

\section{Data Availability}
The data that support the findings of this study are openly available at 
\href{https://github.com/ww1g11/skyrmions-embedded-in-a-vortex}{Github}, Ref.~[\citenum{wang2025a}].

\section{acknowledgments}
We acknowledges financial support from the National Natural Science Foundation of China (Grant No. 12374098).
\bibliographystyle{apsrev4-2}
\bibliography{skyrmion}

\begin{thebibliography}{27}%
\makeatletter
\providecommand \@ifxundefined [1]{%
 \@ifx{#1\undefined}
}%
\providecommand \@ifnum [1]{%
 \ifnum #1\expandafter \@firstoftwo
 \else \expandafter \@secondoftwo
 \fi
}%
\providecommand \@ifx [1]{%
 \ifx #1\expandafter \@firstoftwo
 \else \expandafter \@secondoftwo
 \fi
}%
\providecommand \natexlab [1]{#1}%
\providecommand \enquote  [1]{``#1''}%
\providecommand \bibnamefont  [1]{#1}%
\providecommand \bibfnamefont [1]{#1}%
\providecommand \citenamefont [1]{#1}%
\providecommand \href@noop [0]{\@secondoftwo}%
\providecommand \href [0]{\begingroup \@sanitize@url \@href}%
\providecommand \@href[1]{\@@startlink{#1}\@@href}%
\providecommand \@@href[1]{\endgroup#1\@@endlink}%
\providecommand \@sanitize@url [0]{\catcode `\\12\catcode `\$12\catcode `\&12\catcode `\#12\catcode `\^12\catcode `\_12\catcode `\%12\relax}%
\providecommand \@@startlink[1]{}%
\providecommand \@@endlink[0]{}%
\providecommand \url  [0]{\begingroup\@sanitize@url \@url }%
\providecommand \@url [1]{\endgroup\@href {#1}{\urlprefix }}%
\providecommand \urlprefix  [0]{URL }%
\providecommand \Eprint [0]{\href }%
\providecommand \doibase [0]{https://doi.org/}%
\providecommand \selectlanguage [0]{\@gobble}%
\providecommand \bibinfo  [0]{\@secondoftwo}%
\providecommand \bibfield  [0]{\@secondoftwo}%
\providecommand \translation [1]{[#1]}%
\providecommand \BibitemOpen [0]{}%
\providecommand \bibitemStop [0]{}%
\providecommand \bibitemNoStop [0]{.\EOS\space}%
\providecommand \EOS [0]{\spacefactor3000\relax}%
\providecommand \BibitemShut  [1]{\csname bibitem#1\endcsname}%
\let\auto@bib@innerbib\@empty
\bibitem [{\citenamefont {Shinjo}\ \emph {et~al.}(2000)\citenamefont {Shinjo}, \citenamefont {Okuno}, \citenamefont {Hassdorf}, \citenamefont {Shigeto},\ and\ \citenamefont {Ono}}]{shinjo2000}%
  \BibitemOpen
  \bibfield  {author} {\bibinfo {author} {\bibfnamefont {T.}~\bibnamefont {Shinjo}}, \bibinfo {author} {\bibfnamefont {T.}~\bibnamefont {Okuno}}, \bibinfo {author} {\bibfnamefont {R.}~\bibnamefont {Hassdorf}}, \bibinfo {author} {\bibfnamefont {{\dag}.~K.}\ \bibnamefont {Shigeto}},\ and\ \bibinfo {author} {\bibfnamefont {T.}~\bibnamefont {Ono}},\ }\href {https://doi.org/10.1126/science.289.5481.930} {\bibfield  {journal} {\bibinfo  {journal} {Science}\ }\textbf {\bibinfo {volume} {289}},\ \bibinfo {pages} {930} (\bibinfo {year} {2000})}\BibitemShut {NoStop}%
\bibitem [{\citenamefont {Wachowiak}\ \emph {et~al.}(2002)\citenamefont {Wachowiak}, \citenamefont {Wiebe}, \citenamefont {Bode}, \citenamefont {Pietzsch}, \citenamefont {Morgenstern},\ and\ \citenamefont {Wiesendanger}}]{wachowiak2002}%
  \BibitemOpen
  \bibfield  {author} {\bibinfo {author} {\bibfnamefont {A.}~\bibnamefont {Wachowiak}}, \bibinfo {author} {\bibfnamefont {J.}~\bibnamefont {Wiebe}}, \bibinfo {author} {\bibfnamefont {M.}~\bibnamefont {Bode}}, \bibinfo {author} {\bibfnamefont {O.}~\bibnamefont {Pietzsch}}, \bibinfo {author} {\bibfnamefont {M.}~\bibnamefont {Morgenstern}},\ and\ \bibinfo {author} {\bibfnamefont {R.}~\bibnamefont {Wiesendanger}},\ }\href {https://doi.org/10.1126/science.1075302} {\bibfield  {journal} {\bibinfo  {journal} {Science}\ }\textbf {\bibinfo {volume} {298}},\ \bibinfo {pages} {577} (\bibinfo {year} {2002})}\BibitemShut {NoStop}%
\bibitem [{\citenamefont {Guslienko}\ \emph {et~al.}(2006)\citenamefont {Guslienko}, \citenamefont {Han}, \citenamefont {Keavney}, \citenamefont {Divan},\ and\ \citenamefont {Bader}}]{Guslienko2006c}%
  \BibitemOpen
  \bibfield  {author} {\bibinfo {author} {\bibfnamefont {K.~{\relax Yu}.}\ \bibnamefont {Guslienko}}, \bibinfo {author} {\bibfnamefont {X.~F.}\ \bibnamefont {Han}}, \bibinfo {author} {\bibfnamefont {D.~J.}\ \bibnamefont {Keavney}}, \bibinfo {author} {\bibfnamefont {R.}~\bibnamefont {Divan}},\ and\ \bibinfo {author} {\bibfnamefont {S.~D.}\ \bibnamefont {Bader}},\ }\href {https://doi.org/10.1103/PhysRevLett.96.067205} {\bibfield  {journal} {\bibinfo  {journal} {Physical Review Letters}\ }\textbf {\bibinfo {volume} {96}},\ \bibinfo {pages} {067205} (\bibinfo {year} {2006})}\BibitemShut {NoStop}%
\bibitem [{\citenamefont {Chung}\ \emph {et~al.}(2010)\citenamefont {Chung}, \citenamefont {McMichael}, \citenamefont {Pierce},\ and\ \citenamefont {Unguris}}]{Chung2010}%
  \BibitemOpen
  \bibfield  {author} {\bibinfo {author} {\bibfnamefont {S.-H.}\ \bibnamefont {Chung}}, \bibinfo {author} {\bibfnamefont {R.~D.}\ \bibnamefont {McMichael}}, \bibinfo {author} {\bibfnamefont {D.~T.}\ \bibnamefont {Pierce}},\ and\ \bibinfo {author} {\bibfnamefont {J.}~\bibnamefont {Unguris}},\ }\href {https://doi.org/10.1103/PhysRevB.81.024410} {\bibfield  {journal} {\bibinfo  {journal} {Physical Review B}\ }\textbf {\bibinfo {volume} {81}},\ \bibinfo {pages} {024410} (\bibinfo {year} {2010})}\BibitemShut {NoStop}%
\bibitem [{\citenamefont {Jaafar}\ \emph {et~al.}(2010)\citenamefont {Jaafar}, \citenamefont {Yanes}, \citenamefont {Perez De~Lara}, \citenamefont {{Chubykalo-Fesenko}}, \citenamefont {Asenjo}, \citenamefont {Gonzalez}, \citenamefont {Anguita}, \citenamefont {Vazquez},\ and\ \citenamefont {Vicent}}]{jaafar2010}%
  \BibitemOpen
  \bibfield  {author} {\bibinfo {author} {\bibfnamefont {M.}~\bibnamefont {Jaafar}}, \bibinfo {author} {\bibfnamefont {R.}~\bibnamefont {Yanes}}, \bibinfo {author} {\bibfnamefont {D.}~\bibnamefont {Perez De~Lara}}, \bibinfo {author} {\bibfnamefont {O.}~\bibnamefont {{Chubykalo-Fesenko}}}, \bibinfo {author} {\bibfnamefont {A.}~\bibnamefont {Asenjo}}, \bibinfo {author} {\bibfnamefont {E.~M.}\ \bibnamefont {Gonzalez}}, \bibinfo {author} {\bibfnamefont {J.~V.}\ \bibnamefont {Anguita}}, \bibinfo {author} {\bibfnamefont {M.}~\bibnamefont {Vazquez}},\ and\ \bibinfo {author} {\bibfnamefont {J.~L.}\ \bibnamefont {Vicent}},\ }\href {https://doi.org/10.1103/PhysRevB.81.054439} {\bibfield  {journal} {\bibinfo  {journal} {Physical Review B}\ }\textbf {\bibinfo {volume} {81}},\ \bibinfo {pages} {054439} (\bibinfo {year} {2010})}\BibitemShut {NoStop}%
\bibitem [{\citenamefont {Cambel}\ and\ \citenamefont {Karapetrov}(2011)}]{cambel2011a}%
  \BibitemOpen
  \bibfield  {author} {\bibinfo {author} {\bibfnamefont {V.}~\bibnamefont {Cambel}}\ and\ \bibinfo {author} {\bibfnamefont {G.}~\bibnamefont {Karapetrov}},\ }\href {https://doi.org/10.1103/PhysRevB.84.014424} {\bibfield  {journal} {\bibinfo  {journal} {Physical Review B}\ }\textbf {\bibinfo {volume} {84}},\ \bibinfo {pages} {014424} (\bibinfo {year} {2011})}\BibitemShut {NoStop}%
\bibitem [{\citenamefont {Dzyaloshinskii}(1958)}]{Dzyaloshinskii1958}%
  \BibitemOpen
  \bibfield  {author} {\bibinfo {author} {\bibfnamefont {I.}~\bibnamefont {Dzyaloshinskii}},\ }\href@noop {} {\bibfield  {journal} {\bibinfo  {journal} {J. Phys. Chem. Solids}\ }\textbf {\bibinfo {volume} {4}},\ \bibinfo {pages} {241} (\bibinfo {year} {1958})}\BibitemShut {NoStop}%
\bibitem [{\citenamefont {Moriya}(1960)}]{Moriya1960}%
  \BibitemOpen
  \bibfield  {author} {\bibinfo {author} {\bibfnamefont {T.}~\bibnamefont {Moriya}},\ }\href@noop {} {\bibfield  {journal} {\bibinfo  {journal} {Phys. Rev.}\ }\textbf {\bibinfo {volume} {120}},\ \bibinfo {pages} {91} (\bibinfo {year} {1960})}\BibitemShut {NoStop}%
\bibitem [{\citenamefont {Bogdanov}\ and\ \citenamefont {R{\"o}{\ss}ler}(2001)}]{bogdanov2001}%
  \BibitemOpen
  \bibfield  {author} {\bibinfo {author} {\bibfnamefont {A.}~\bibnamefont {Bogdanov}}\ and\ \bibinfo {author} {\bibfnamefont {U.}~\bibnamefont {R{\"o}{\ss}ler}},\ }\href {https://doi.org/10.1103/PhysRevLett.87.037203} {\bibfield  {journal} {\bibinfo  {journal} {Physical Review Letters}\ }\textbf {\bibinfo {volume} {87}},\ \bibinfo {pages} {037203} (\bibinfo {year} {2001})}\BibitemShut {NoStop}%
\bibitem [{\citenamefont {Rohart}\ and\ \citenamefont {Thiaville}(2013)}]{rohart2013}%
  \BibitemOpen
  \bibfield  {author} {\bibinfo {author} {\bibfnamefont {S.}~\bibnamefont {Rohart}}\ and\ \bibinfo {author} {\bibfnamefont {A.}~\bibnamefont {Thiaville}},\ }\href {https://doi.org/10.1103/PhysRevB.88.184422} {\bibfield  {journal} {\bibinfo  {journal} {Physical Review B}\ }\textbf {\bibinfo {volume} {88}},\ \bibinfo {pages} {184422} (\bibinfo {year} {2013})},\ \Eprint {https://arxiv.org/abs/1310.0666} {arXiv:1310.0666} \BibitemShut {NoStop}%
\bibitem [{\citenamefont {{Moreau-Luchaire}}\ \emph {et~al.}(2016)\citenamefont {{Moreau-Luchaire}}, \citenamefont {Moutafis}, \citenamefont {Reyren}, \citenamefont {Sampaio}, \citenamefont {Vaz}, \citenamefont {Van~Horne}, \citenamefont {Bouzehouane}, \citenamefont {Garcia}, \citenamefont {Deranlot}, \citenamefont {Warnicke}, \citenamefont {Wohlh{\"u}ter}, \citenamefont {George}, \citenamefont {Weigand}, \citenamefont {Raabe}, \citenamefont {Cros},\ and\ \citenamefont {Fert}}]{moreau-luchaire2016}%
  \BibitemOpen
  \bibfield  {author} {\bibinfo {author} {\bibfnamefont {C.}~\bibnamefont {{Moreau-Luchaire}}}, \bibinfo {author} {\bibfnamefont {C.}~\bibnamefont {Moutafis}}, \bibinfo {author} {\bibfnamefont {N.}~\bibnamefont {Reyren}}, \bibinfo {author} {\bibfnamefont {J.}~\bibnamefont {Sampaio}}, \bibinfo {author} {\bibfnamefont {C.~A.~F.}\ \bibnamefont {Vaz}}, \bibinfo {author} {\bibfnamefont {N.}~\bibnamefont {Van~Horne}}, \bibinfo {author} {\bibfnamefont {K.}~\bibnamefont {Bouzehouane}}, \bibinfo {author} {\bibfnamefont {K.}~\bibnamefont {Garcia}}, \bibinfo {author} {\bibfnamefont {C.}~\bibnamefont {Deranlot}}, \bibinfo {author} {\bibfnamefont {P.}~\bibnamefont {Warnicke}}, \bibinfo {author} {\bibfnamefont {P.}~\bibnamefont {Wohlh{\"u}ter}}, \bibinfo {author} {\bibfnamefont {J.-M.}\ \bibnamefont {George}}, \bibinfo {author} {\bibfnamefont {M.}~\bibnamefont {Weigand}}, \bibinfo {author} {\bibfnamefont {J.}~\bibnamefont {Raabe}}, \bibinfo {author} {\bibfnamefont {V.}~\bibnamefont {Cros}},\ and\ \bibinfo {author} {\bibfnamefont {A.}~\bibnamefont {Fert}},\ }\href {https://doi.org/10.1038/nnano.2015.313} {\bibfield  {journal} {\bibinfo  {journal} {Nature Nanotechnology}\ }\textbf {\bibinfo {volume} {11}},\ \bibinfo {pages} {444} (\bibinfo {year} {2016})}\BibitemShut {NoStop}%
\bibitem [{\citenamefont {Zhang}\ \emph {et~al.}(2015)\citenamefont {Zhang}, \citenamefont {Wang}, \citenamefont {Beg}, \citenamefont {Fangohr},\ and\ \citenamefont {Kuch}}]{Zhang2015}%
  \BibitemOpen
  \bibfield  {author} {\bibinfo {author} {\bibfnamefont {B.}~\bibnamefont {Zhang}}, \bibinfo {author} {\bibfnamefont {W.}~\bibnamefont {Wang}}, \bibinfo {author} {\bibfnamefont {M.}~\bibnamefont {Beg}}, \bibinfo {author} {\bibfnamefont {H.}~\bibnamefont {Fangohr}},\ and\ \bibinfo {author} {\bibfnamefont {W.}~\bibnamefont {Kuch}},\ }\href {https://doi.org/10.1063/1.4914496} {\bibfield  {journal} {\bibinfo  {journal} {Applied Physics Letters}\ }\textbf {\bibinfo {volume} {106}},\ \bibinfo {pages} {102401} (\bibinfo {year} {2015})},\ \Eprint {https://arxiv.org/abs/1503.02869} {arXiv:1503.02869} \BibitemShut {NoStop}%
\bibitem [{\citenamefont {Fert}\ \emph {et~al.}(2017)\citenamefont {Fert}, \citenamefont {Reyren},\ and\ \citenamefont {Cros}}]{Fert2017}%
  \BibitemOpen
  \bibfield  {author} {\bibinfo {author} {\bibfnamefont {A.}~\bibnamefont {Fert}}, \bibinfo {author} {\bibfnamefont {N.}~\bibnamefont {Reyren}},\ and\ \bibinfo {author} {\bibfnamefont {V.}~\bibnamefont {Cros}},\ }\href {https://doi.org/10.1038/natrevmats.2017.31} {\bibfield  {journal} {\bibinfo  {journal} {Nature Reviews Materials}\ }\textbf {\bibinfo {volume} {2}},\ \bibinfo {pages} {17031} (\bibinfo {year} {2017})}\BibitemShut {NoStop}%
\bibitem [{\citenamefont {Siracusano}\ \emph {et~al.}(2016)\citenamefont {Siracusano}, \citenamefont {Tomasello}, \citenamefont {Giordano}, \citenamefont {Puliafito}, \citenamefont {Azzerboni}, \citenamefont {Ozatay}, \citenamefont {Carpentieri},\ and\ \citenamefont {Finocchio}}]{Siracusano2016b}%
  \BibitemOpen
  \bibfield  {author} {\bibinfo {author} {\bibfnamefont {G.}~\bibnamefont {Siracusano}}, \bibinfo {author} {\bibfnamefont {R.}~\bibnamefont {Tomasello}}, \bibinfo {author} {\bibfnamefont {A.}~\bibnamefont {Giordano}}, \bibinfo {author} {\bibfnamefont {V.}~\bibnamefont {Puliafito}}, \bibinfo {author} {\bibfnamefont {B.}~\bibnamefont {Azzerboni}}, \bibinfo {author} {\bibfnamefont {O.}~\bibnamefont {Ozatay}}, \bibinfo {author} {\bibfnamefont {M.}~\bibnamefont {Carpentieri}},\ and\ \bibinfo {author} {\bibfnamefont {G.}~\bibnamefont {Finocchio}},\ }\href {https://doi.org/10.1103/PhysRevLett.117.087204} {\bibfield  {journal} {\bibinfo  {journal} {Physical Review Letters}\ }\textbf {\bibinfo {volume} {117}},\ \bibinfo {pages} {087204} (\bibinfo {year} {2016})}\BibitemShut {NoStop}%
\bibitem [{\citenamefont {Van~Waeyenberge}\ \emph {et~al.}(2006)\citenamefont {Van~Waeyenberge}, \citenamefont {Puzic}, \citenamefont {Stoll}, \citenamefont {Chou}, \citenamefont {Tyliszczak}, \citenamefont {Hertel}, \citenamefont {F{\"a}hnle}, \citenamefont {Br{\"u}ckl}, \citenamefont {Rott}, \citenamefont {Reiss}, \citenamefont {Neudecker}, \citenamefont {Weiss}, \citenamefont {Back},\ and\ \citenamefont {Sch{\"u}tz}}]{vanwaeyenberge2006a}%
  \BibitemOpen
  \bibfield  {author} {\bibinfo {author} {\bibfnamefont {B.}~\bibnamefont {Van~Waeyenberge}}, \bibinfo {author} {\bibfnamefont {A.}~\bibnamefont {Puzic}}, \bibinfo {author} {\bibfnamefont {H.}~\bibnamefont {Stoll}}, \bibinfo {author} {\bibfnamefont {K.~W.}\ \bibnamefont {Chou}}, \bibinfo {author} {\bibfnamefont {T.}~\bibnamefont {Tyliszczak}}, \bibinfo {author} {\bibfnamefont {R.}~\bibnamefont {Hertel}}, \bibinfo {author} {\bibfnamefont {M.}~\bibnamefont {F{\"a}hnle}}, \bibinfo {author} {\bibfnamefont {H.}~\bibnamefont {Br{\"u}ckl}}, \bibinfo {author} {\bibfnamefont {K.}~\bibnamefont {Rott}}, \bibinfo {author} {\bibfnamefont {G.}~\bibnamefont {Reiss}}, \bibinfo {author} {\bibfnamefont {I.}~\bibnamefont {Neudecker}}, \bibinfo {author} {\bibfnamefont {D.}~\bibnamefont {Weiss}}, \bibinfo {author} {\bibfnamefont {C.~H.}\ \bibnamefont {Back}},\ and\ \bibinfo {author} {\bibfnamefont {G.}~\bibnamefont {Sch{\"u}tz}},\ }\href {https://doi.org/10.1038/nature05240} {\bibfield  {journal} {\bibinfo  {journal} {Nature}\ }\textbf {\bibinfo {volume} {444}},\ \bibinfo {pages} {461} (\bibinfo {year} {2006})}\BibitemShut {NoStop}%
\bibitem [{\citenamefont {Hertel}\ \emph {et~al.}(2007)\citenamefont {Hertel}, \citenamefont {Gliga}, \citenamefont {F{\"a}hnle},\ and\ \citenamefont {Schneider}}]{hertel2007a}%
  \BibitemOpen
  \bibfield  {author} {\bibinfo {author} {\bibfnamefont {R.}~\bibnamefont {Hertel}}, \bibinfo {author} {\bibfnamefont {S.}~\bibnamefont {Gliga}}, \bibinfo {author} {\bibfnamefont {M.}~\bibnamefont {F{\"a}hnle}},\ and\ \bibinfo {author} {\bibfnamefont {C.~M.}\ \bibnamefont {Schneider}},\ }\href {https://doi.org/10.1103/PhysRevLett.98.117201} {\bibfield  {journal} {\bibinfo  {journal} {Physical Review Letters}\ }\textbf {\bibinfo {volume} {98}},\ \bibinfo {pages} {117201} (\bibinfo {year} {2007})}\BibitemShut {NoStop}%
\bibitem [{\citenamefont {Wang}\ \emph {et~al.}(2024)\citenamefont {Wang}, \citenamefont {Lyu}, \citenamefont {Kong}, \citenamefont {Fangohr},\ and\ \citenamefont {Du}}]{Wang2024a}%
  \BibitemOpen
  \bibfield  {author} {\bibinfo {author} {\bibfnamefont {W.}~\bibnamefont {Wang}}, \bibinfo {author} {\bibfnamefont {B.}~\bibnamefont {Lyu}}, \bibinfo {author} {\bibfnamefont {L.}~\bibnamefont {Kong}}, \bibinfo {author} {\bibfnamefont {H.}~\bibnamefont {Fangohr}},\ and\ \bibinfo {author} {\bibfnamefont {H.}~\bibnamefont {Du}},\ }\href {https://doi.org/10.1088/1674-1056/ad766f} {\bibfield  {journal} {\bibinfo  {journal} {Chinese Physics B}\ }\textbf {\bibinfo {volume} {33}},\ \bibinfo {pages} {107508} (\bibinfo {year} {2024})}\BibitemShut {NoStop}%
\bibitem [{\citenamefont {Mustaghfiroh}\ \emph {et~al.}(2019)\citenamefont {Mustaghfiroh}, \citenamefont {Kurniawan},\ and\ \citenamefont {Djuhana}}]{Mustaghfiroh2019}%
  \BibitemOpen
  \bibfield  {author} {\bibinfo {author} {\bibfnamefont {Q.}~\bibnamefont {Mustaghfiroh}}, \bibinfo {author} {\bibfnamefont {C.}~\bibnamefont {Kurniawan}},\ and\ \bibinfo {author} {\bibfnamefont {D.}~\bibnamefont {Djuhana}},\ }\href {https://doi.org/10.1088/1757-899X/553/1/012009} {\bibfield  {journal} {\bibinfo  {journal} {IOP Conference Series: Materials Science and Engineering}\ }\textbf {\bibinfo {volume} {553}},\ \bibinfo {pages} {012009} (\bibinfo {year} {2019})}\BibitemShut {NoStop}%
\bibitem [{\citenamefont {Nagaosa}\ and\ \citenamefont {Tokura}(2013)}]{Nagaosa2013}%
  \BibitemOpen
  \bibfield  {author} {\bibinfo {author} {\bibfnamefont {N.}~\bibnamefont {Nagaosa}}\ and\ \bibinfo {author} {\bibfnamefont {Y.}~\bibnamefont {Tokura}},\ }\href {https://doi.org/10.1038/nnano.2013.243} {\bibfield  {journal} {\bibinfo  {journal} {Nature Nanotechnology}\ }\textbf {\bibinfo {volume} {8}},\ \bibinfo {pages} {899} (\bibinfo {year} {2013})}\BibitemShut {NoStop}%
\bibitem [{\citenamefont {Lin}\ \emph {et~al.}(2015)\citenamefont {Lin}, \citenamefont {Saxena},\ and\ \citenamefont {Batista}}]{Lin2015}%
  \BibitemOpen
  \bibfield  {author} {\bibinfo {author} {\bibfnamefont {S.-Z.}\ \bibnamefont {Lin}}, \bibinfo {author} {\bibfnamefont {A.}~\bibnamefont {Saxena}},\ and\ \bibinfo {author} {\bibfnamefont {C.~D.}\ \bibnamefont {Batista}},\ }\href {https://doi.org/10.1103/PhysRevB.91.224407} {\bibfield  {journal} {\bibinfo  {journal} {Physical Review B}\ }\textbf {\bibinfo {volume} {91}},\ \bibinfo {pages} {224407} (\bibinfo {year} {2015})}\BibitemShut {NoStop}%
\bibitem [{\citenamefont {Yu}\ \emph {et~al.}(2018)\citenamefont {Yu}, \citenamefont {Koshibae}, \citenamefont {Tokunaga}, \citenamefont {Shibata}, \citenamefont {Taguchi}, \citenamefont {Nagaosa},\ and\ \citenamefont {Tokura}}]{Yu2018}%
  \BibitemOpen
  \bibfield  {author} {\bibinfo {author} {\bibfnamefont {X.~Z.}\ \bibnamefont {Yu}}, \bibinfo {author} {\bibfnamefont {W.}~\bibnamefont {Koshibae}}, \bibinfo {author} {\bibfnamefont {Y.}~\bibnamefont {Tokunaga}}, \bibinfo {author} {\bibfnamefont {K.}~\bibnamefont {Shibata}}, \bibinfo {author} {\bibfnamefont {Y.}~\bibnamefont {Taguchi}}, \bibinfo {author} {\bibfnamefont {N.}~\bibnamefont {Nagaosa}},\ and\ \bibinfo {author} {\bibfnamefont {Y.}~\bibnamefont {Tokura}},\ }\href {https://doi.org/10.1038/s41586-018-0745-3} {\bibfield  {journal} {\bibinfo  {journal} {Nature}\ }\textbf {\bibinfo {volume} {564}},\ \bibinfo {pages} {95} (\bibinfo {year} {2018})}\BibitemShut {NoStop}%
\bibitem [{\citenamefont {Xia}\ \emph {et~al.}(2022)\citenamefont {Xia}, \citenamefont {Zhang}, \citenamefont {Liu}, \citenamefont {Zhou},\ and\ \citenamefont {Ezawa}}]{Xia2022}%
  \BibitemOpen
  \bibfield  {author} {\bibinfo {author} {\bibfnamefont {J.}~\bibnamefont {Xia}}, \bibinfo {author} {\bibfnamefont {X.}~\bibnamefont {Zhang}}, \bibinfo {author} {\bibfnamefont {X.}~\bibnamefont {Liu}}, \bibinfo {author} {\bibfnamefont {Y.}~\bibnamefont {Zhou}},\ and\ \bibinfo {author} {\bibfnamefont {M.}~\bibnamefont {Ezawa}},\ }\href {https://doi.org/10.1038/s43246-022-00311-w} {\bibfield  {journal} {\bibinfo  {journal} {Communications Materials}\ }\textbf {\bibinfo {volume} {3}},\ \bibinfo {pages} {88} (\bibinfo {year} {2022})}\BibitemShut {NoStop}%
\bibitem [{\citenamefont {Foster}\ \emph {et~al.}(2019)\citenamefont {Foster}, \citenamefont {Kind}, \citenamefont {Ackerman}, \citenamefont {Tai}, \citenamefont {Dennis},\ and\ \citenamefont {Smalyukh}}]{Foster2019}%
  \BibitemOpen
  \bibfield  {author} {\bibinfo {author} {\bibfnamefont {D.}~\bibnamefont {Foster}}, \bibinfo {author} {\bibfnamefont {C.}~\bibnamefont {Kind}}, \bibinfo {author} {\bibfnamefont {P.~J.}\ \bibnamefont {Ackerman}}, \bibinfo {author} {\bibfnamefont {J.-S.~B.}\ \bibnamefont {Tai}}, \bibinfo {author} {\bibfnamefont {M.~R.}\ \bibnamefont {Dennis}},\ and\ \bibinfo {author} {\bibfnamefont {I.~I.}\ \bibnamefont {Smalyukh}},\ }\href {https://doi.org/10/gfxrhv} {\bibfield  {journal} {\bibinfo  {journal} {Nature Physics}\ }\textbf {\bibinfo {volume} {15}},\ \bibinfo {pages} {655} (\bibinfo {year} {2019})}\BibitemShut {NoStop}%
\bibitem [{\citenamefont {Kind}\ and\ \citenamefont {Foster}(2021)}]{Kind2021}%
  \BibitemOpen
  \bibfield  {author} {\bibinfo {author} {\bibfnamefont {C.}~\bibnamefont {Kind}}\ and\ \bibinfo {author} {\bibfnamefont {D.}~\bibnamefont {Foster}},\ }\href {https://doi.org/10.1103/PhysRevB.103.L100413} {\bibfield  {journal} {\bibinfo  {journal} {Physical Review B}\ }\textbf {\bibinfo {volume} {103}},\ \bibinfo {pages} {L100413} (\bibinfo {year} {2021})}\BibitemShut {NoStop}%
\bibitem [{\citenamefont {Rybakov}\ and\ \citenamefont {Kiselev}(2019)}]{Rybakov2019}%
  \BibitemOpen
  \bibfield  {author} {\bibinfo {author} {\bibfnamefont {F.~N.}\ \bibnamefont {Rybakov}}\ and\ \bibinfo {author} {\bibfnamefont {N.~S.}\ \bibnamefont {Kiselev}},\ }\href {https://doi.org/10.1103/PhysRevB.99.064437} {\bibfield  {journal} {\bibinfo  {journal} {Physical Review B}\ }\textbf {\bibinfo {volume} {99}},\ \bibinfo {pages} {064437} (\bibinfo {year} {2019})}\BibitemShut {NoStop}%
\bibitem [{\citenamefont {Banerjee}\ \emph {et~al.}(2014)\citenamefont {Banerjee}, \citenamefont {Rowland}, \citenamefont {Erten},\ and\ \citenamefont {Randeria}}]{Banerjee2014}%
  \BibitemOpen
  \bibfield  {author} {\bibinfo {author} {\bibfnamefont {S.}~\bibnamefont {Banerjee}}, \bibinfo {author} {\bibfnamefont {J.}~\bibnamefont {Rowland}}, \bibinfo {author} {\bibfnamefont {O.}~\bibnamefont {Erten}},\ and\ \bibinfo {author} {\bibfnamefont {M.}~\bibnamefont {Randeria}},\ }\href {https://doi.org/10.1103/PhysRevX.4.031045} {\bibfield  {journal} {\bibinfo  {journal} {Physical Review X}\ }\textbf {\bibinfo {volume} {4}},\ \bibinfo {pages} {031045} (\bibinfo {year} {2014})}\BibitemShut {NoStop}%
\bibitem [{\citenamefont {Wang}(2024)}]{wang2025a}%
  \BibitemOpen
  \bibfield  {author} {\bibinfo {author} {\bibfnamefont {W.}~\bibnamefont {Wang}},\ }\href {https://github.com/ww1g11/skyrmions-embedded-in-a-vortex} {\bibinfo {title} {Code and data for "magnetic skyrmions embedded in a vortex"}} (\bibinfo {year} {2024})\BibitemShut {NoStop}%
\end{thebibliography}%

\end{document}